\documentclass[prb,aps,twocolumn,dcolumn,superscriptaddress,amsmath,amssymb]{revtex4}

\usepackage{graphicx}
\usepackage{graphicx}
\begin{document}

\title{Comment on ``Hole digging in ensembles of tunneling molecular magnets''}
\author{Juan J. Alonso}
\affiliation{F\'{\i}sica Aplicada I, Universidad de M\'alaga,
29071-M\'alaga, Spain}
\email[E-mail address: ] {jjalonso@Darnitsa.Cie.Uma.Es}
\author{Julio F. Fern\'andez}
\affiliation{ICMA, CSIC and Universidad de Zaragoza, 50009-Zaragoza, Spain}
\email[E-mail address: ] {JFF@Pipe.Unizar.Es}
\homepage[ URL: ] {http://Pipe.Unizar.Es/~jff}
\date{\today}


\begin{abstract}

Tupitsyn et al. [Phys. Rev. B {\bf 69}, 132406 (2004)]
have recently reported results for the relaxation of
crystalline systems of single--molecule magnets, such as Fe$_8 $.
They claim that, quite generally,              
(1) the magnetization and 
hole widths of field--distributions evolve with 
time $t$ as $ \sqrt{t}$, and (2) the holes' line shapes are
Lorentzian.
We give a counter--example to these conclusions,
and show that the main assumption on which some of them rest
is invalid.

\end{abstract}

\maketitle

\begin{figure}[!ht]
\includegraphics*[width=80mm]{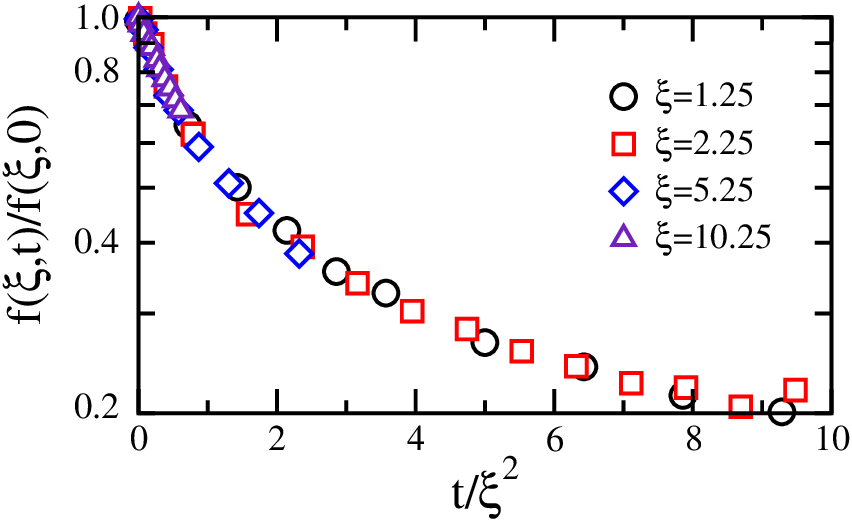}
\caption{The field function $f(\xi ,t)$ versus
$t/\xi^2$ for SC lattices and the
shown values of the dipole field $\xi$.
The tunnel window's half--width $\xi_0$ is $0.1$.
The initial value of the magnetization $m_0$
is 0.2 of saturation.
All data points are for systems of 32768 spins,
and follow from averages over
$ 800$ runs,
over times much greater than
$t\gg \tau_0$ but within the range where $m\propto t^p$.}
\label{nvst1}
\end{figure}

Tupitsyn, Stamp, and Prokof'ev\cite{TPS} (TSP)
have recently reported results for the relaxation of
crystalline systems of single--molecule magnets, such as Fe$_8 $,
whose tunneling windows are much narrower than their dipolar field spread.
They find that, independently
of crystalline lattice structure, (1) the magnetization $m$ and
hole widths of field--distributions evolve with
time $t$ as $ \sqrt{t}$, and (2) field--distribution line shapes are
Lorentzian.
The first conclusion is contrary to our prediction,
of Ref. [\onlinecite{ourPRB}],
that, in zero field,
$m$ relaxes from weakly polarized states
as $ t^p$, where
$p$ depends on crystal structure.
We give below a counter--example to the conclusions TSP have reached,
and show that the main assumption on which some of their conclusions rest
is incorrect.

\begin{figure}[!ht]
\includegraphics*[width=80mm]{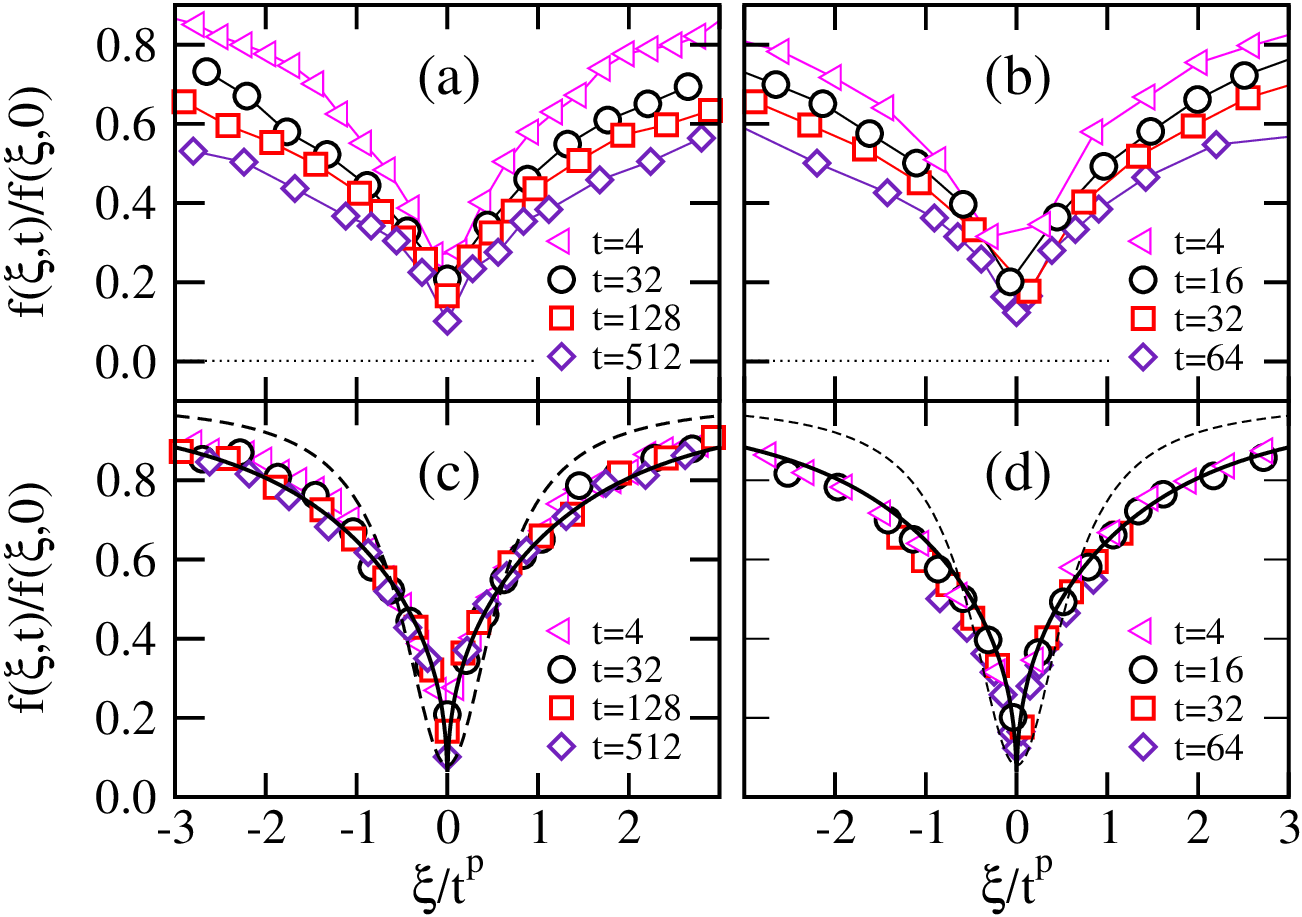}
\caption{(a) $f(\xi ,t)$ versus $\xi/t^p$ for
FCC lattices and the shown values
of $t$,
$\xi_0=0.01$, and $p=0.5$. Lines are guides to the eye.
(b) Same as in (a) but for $\xi_0=0.1$. (c) Same
as in (a) but for $p=0.726$. The dashed line is for the
best fitting Lorentz curve. The full line is from our theory.
(d) Same as in (c) but for $\xi_0=0.1$.
In all cases, $m_0=0.2$ 
All data points are for systems of 65536 spins, and follow from averages over
$ 800$ runs over times much greater than
$t\gg \tau_0$ but within the range where $m\propto t^p$.
All the shown data points
come from the $2\xi_0<\xi<0.2\delta \xi$ (nearly one and two decades
for $\xi_0=0.1$ and
$\xi_0=0.01$, respectively).
In the shown range  
of time values, 
$1-m/m_0$
changes by approximately one and two
decades (see Fig. 1), for $\xi_0=0.1$ and $\xi_0=0.01$, respectively.}
\label{nvst2}
\end{figure}

We mainly use the notation of Ref. [\onlinecite{TPS}], giving the
bias field $\xi$ in
terms of the tunnel window field $\xi_0$, but $\xi_0$ is given in terms of 
the nearest neighbor dipolar field $E_D$. 
We assume
spins flip at rate $1/\tau_0$ if
$\mid \xi \mid <1$, but not at all otherwise, and
time $t$ is given in terms of $\tau_0$.
The following numbers
may be found useful: the rms value of the dipolar field $\delta \xi$ is
$3.7E_D$ and $8.3E_D$ for SC and FCC, respectively. 

Let $p_\uparrow (\xi ,t)$ 
[$p_\downarrow (\xi ,t)$] be the number density 
of up--spins (down--spins) with a field $\xi$ 
acting on them, and let 
$f(\xi ,t)=p_\downarrow (\xi ,t)-p_\uparrow (\xi ,t)$.
Note that $m(t)=-\int d\xi f(\xi ,t)$.
The main ingredient underlying
 Eq. (1) of Ref. 
[\onlinecite{TPS}] is the assumption 
that $f(\xi ,t)\propto N(\xi )\exp [-t/\tau (\xi )]$, where $N(\xi )$ is 
of no interest to us here,
and $\tau (\xi )$ is some time 
that depends only on $\xi $. The Monte Carlo (MC) 
results shown in Fig. 1 are
contrary to the assumption of TSP, that $f(\xi ,t)$ is 
exponential in $t$.\cite{razon}
[The probability density
that a spin have field $\xi$ 
and has not yet flipped at time ``t''
behaves much as $f(\xi ,t)$.]

From the assumption that $f(\xi ,t)\propto \exp [-t/\tau (\xi )]$ and 
the further general statement TSP make,
that $1/\tau (\xi )$ is a Lorentzian function of $\xi$,
hole line widths 
that grow as $\sqrt{t}$ when $t\gtrsim \tau_0$ 
follow in Ref. [\onlinecite{TPS}].
Such $\sqrt{t}$ growth does take place in SC lattices, but
not in general, as one can gather
from the MC results exhibited
in Figs. 2a and 2b for FCC lattices.
Rather, from Figs. 2c and 2d, 
one gathers that $\xi$ then scales as $t^p$, 
where 
$p\simeq 0.73$.\cite{penultima}
Finally, in the region of interest, 
when $t\gg \tau_0$ but $t$ is still within 
the range where $m\propto t^p$,
the line shape in Figs. 2c and 2d differ 
significantly from a Lorentzian function. 
For comparison, results from our own 
theory \cite{own} are also shown in Figs. 2c and 2d.
\begin{figure}[!ht]
\includegraphics*[width=80mm]{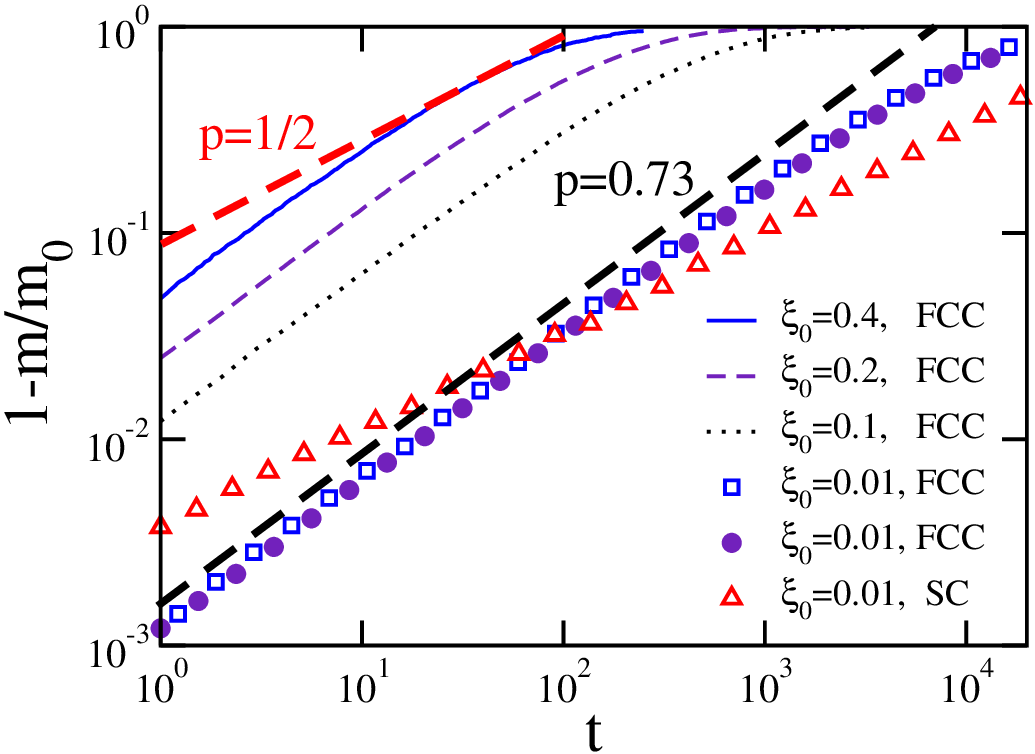}
\caption{$1-m/m_0$ versus time $t$. All data points
follow from MC simulations.
Triangles are for SC lattices. All other data
points are for FCC lattices.
All data points for the FCC lattice are for systems of 8192 spins,
except for $\bullet$, which stand for 65536 spins.
Data points for SC lattices are for 4096 spins.
For all data points, averages over 1200 runs were performed,
except for the 65536 spin system and $t>1200$. For them, averages
over 100 runs were performed.}
\end{figure}

For completeness sake, 
we examine further numerical evidence that supports our claim
that $p=0.73$, not $1/2$, for FCC lattices.
To this end, note first that $\xi\sim t^p$ scaling
implies, through the relation $m(t)=-\int d\xi f(\xi ,t)$,
that $m\sim t^p$.
Thus, the value of
$p$ can also be obtained from the time evolution of $m$.

The difference between the relaxation of
the magnetization in SC and FCC lattices
can be clearly appreciated in Fig. 3,
as well as in Figs. 4a and 4b.
Note in Fig. 3 that, for FCC lattices, 
$m\propto t^p$ and $p\simeq 0.73$
for time spans that are increasingly
larger for larger values of $1 /\xi_0$.
(This is as predicted in Ref. [\onlinecite{ourPRB}].)
Note also that the $p=1/2$ slope that
is claimed by TSP to hold universally
for all lattices 
appears to ensue for FCC
lattices only when the relaxation crosses over
from the $m\propto t^p$ regime to saturation.
This is in clear contrast with the data
points for SC lattices in Fig. 3, for which $p\simeq 0.5$.
Linear $m/m_0$ versus time plots for SC and FCC
lattices which further illustrate this point
are also shown in Figs. 4a and 4b.
The small neighborhood of the inflection point 
(marked as a straight line segment) in Fig. 4b
is, of course, nearly straight. This transient
behavior might
be misinterpreted
as the onset of a $m\propto \sqrt{t}$ regime 
if, as in Fig. 1 of Ref. [\onlinecite{arxiv}],
data points below $m/m_0\simeq 0.5$ are not included
in the plot.
\begin{figure}[!ht]
\includegraphics*[width=80mm]{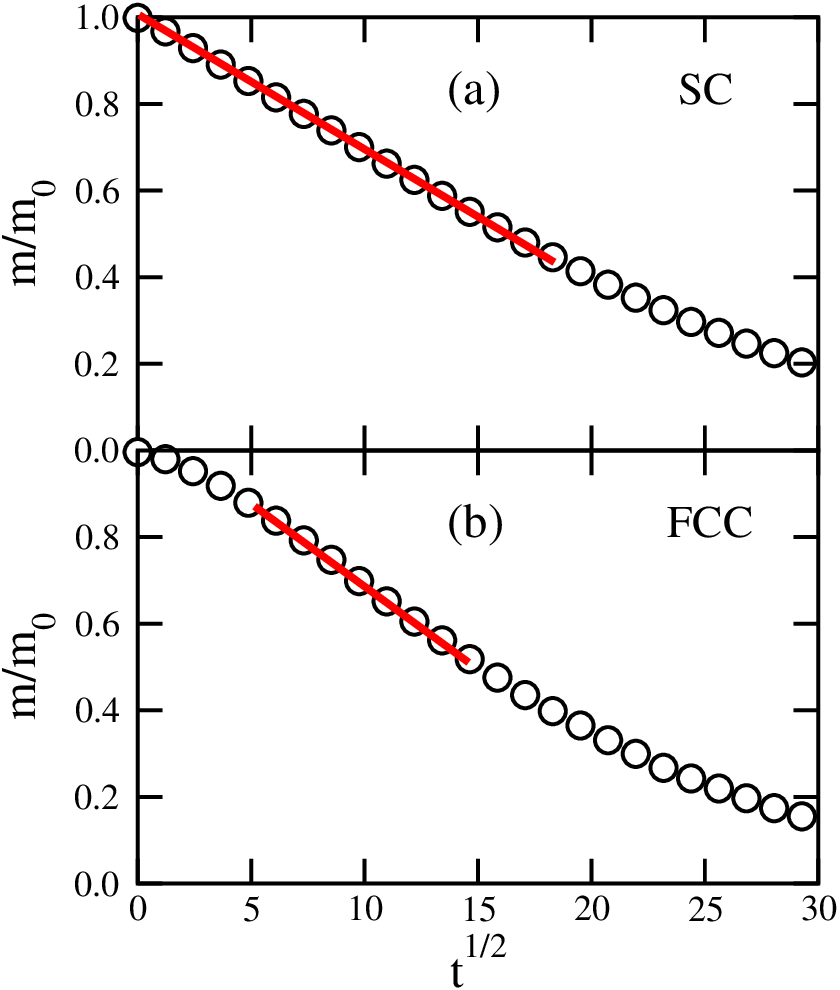}
\caption{(a) $m/m_0$
versus $\sqrt{t}$. Data points are
from averaging over at least 4000 MC 
runs for 4096 spins on SC lattices with $\xi_0=0.1$.
The straight line is a guide to the eye.
(b) Same as in (a) but for 8192 spins on an FCC lattice.
The straight line segment covers a neighborhood of the inflection point.}
\end{figure}

Finally, comparison of the results shown in 
Figs. 2a and 2b, Figs. 2c and 2d, as well as among
different curves shown in Fig. 3 for different values
of $\xi_0$ should allay
any misgivings about spurious effects
that might arise from the finite 
number of spins $n_s$ in the tunnel window,
since $n_s\propto \xi_0$ if
$\xi_0\ll \delta \xi$.
To the same end, data points for two FCC lattice sizes
are shown in Fig. 3 for $\xi_0=0.01$.

Grant BFM2003-03919-C02/FISI, from Ministerio de Ciencia y 
Tecnolog\'{\i}a of Spain, is gratefully
acknowledged.


\begin{thebibliography}{99}
\bibitem{razon}This is because
$\partial f(\xi ,t)/\partial t\neq -f(h,t)/\tau(\xi )$,
since
$\partial f(\xi ,t)/\partial t$ depends, not only on $f$ and $\xi$,
 but also on the rate of magnetization reversal
which is going on within the tunneling window at time $t$, which in turn
depends non--trivially on time.
\end{thebibliography}
\end{document}